\documentclass[aps,prc,twocolumn,superscriptaddress,showpacs]{revtex4}
\usepackage{graphicx}
\usepackage{dcolumn}
\usepackage{bm}
\usepackage{subfigure}
\usepackage{amssymb}
\newcommand{\beq}{\begin{equation}}
\newcommand{\eeq}{\end{equation}}
\newcommand{\beqar}{\begin{eqnarray}}
\newcommand{\eeqar}{\end{eqnarray}}
\newcommand{\ds}{\displaystyle}
\begin{document}

\title{Effect of jets on $v_4/v_2^2$ ratio and constituent quark
scaling in relativistic heavy-ion collisions}

\author{L.~Bravina}
\affiliation{
Department of Physics, University of Oslo, PB 1048 Blindern,
N-0316 Oslo, Norway
\vspace*{1ex}}
\author{B.H.~Brusheim~Johansson}
\affiliation{
Department of Physics, University of Oslo, PB 1048 Blindern,
N-0316 Oslo, Norway
\vspace*{1ex}}
\author{G.~Eyyubova}
\altaffiliation[Also at ]{
Skobeltsyn Institute of Nuclear Physics,
Moscow State University, RU-119991 Moscow, Russia
\vspace*{1ex}}
\affiliation{
Department of Physics, University of Oslo, PB 1048 Blindern,
N-0316 Oslo, Norway
\vspace*{1ex}}
\author{E.~Zabrodin}
\altaffiliation[Also at ]{
Skobeltsyn Institute of Nuclear Physics,
Moscow State University, RU-119991 Moscow, Russia
\vspace*{1ex}}
\affiliation{
Department of Physics, University of Oslo, PB 1048 Blindern,
N-0316 Oslo, Norway
\vspace*{1ex}}

\date{\today}

\begin{abstract}
The Monte Carlo HYDJET++ model, that combines parametrized 
hydrodynamics with jets, is employed to study formation of second 
$v_2$ and fourth $v_4$ components of the anisotropic flow in
ultrarelativistic heavy-ion collisions at energies of the Relativistic 
Heavy Ion Collider (RHIC) and the Large Hadron Collider (LHC),
$\sqrt{s}=200${\it A}~GeV and $\sqrt{s}=2.76${\it A}~TeV, respectively. 
It is shown that the quenched jets contribute to the soft part of the 
$v_2(p_T)$ and $v_4(p_T)$ spectra. The jets increase the ratio 
$v_4/v_2^2$ thus leading to deviations of the ratio from the value of 
0.5 predicted by the ideal hydrodynamics. Together with the 
event-by-event fluctuations, the influence of jets can explain 
quantitatively the ratio $v_4/v_2^2$ at $p_T \leq 2$\,GeV/$c$ for both
energies and qualitatively the rise of its high-$p_T$ tail at LHC. Jets 
are also responsible for violation of the number-of-constituent-quark 
(NCQ) scaling at LHC despite the fact that the scaling is 
fulfilled for the hydro- part of particle spectra.
\end{abstract}


\maketitle
\section{Introduction}
\label{sec1}

The transverse collective flow of particles is an important 
characteristic of ultrarelativistic heavy-ion collisions because the 
flow is able to carry information about the early stage of the 
reaction. Particularly, the collective flow is very sensitive to 
change of the equation of state (EOS), e.g., during the quark-hadron 
phase transition. The azimuthal distribution of particles can be cast 
\cite{VoZh96,PoVo98} in the form of Fourier series
\beq
\ds
E \frac{d^3 N}{d^3 p} = \frac{1}{\pi} \frac{d^2 N}{dp_t^2 dy} \left[
1 + \sum_{n=1}^{\infty} 2 v_n \cos(n\phi) \right] .
\label{eq1}
\eeq
Here $\phi$, $p_t$, and $y$ are the azimuthal angle, the transverse 
momentum, and the rapidity of a particle, respectively. The unity in 
the parentheses represents the isotropic radial flow, whereas the sum
of harmonics refers to anisotropic flow. The first two harmonics of 
the anisotropic flow, dubbed directed flow $v_1$ and elliptic 
flow $v_2$, have been extensively studied both experimentally and 
theoretically in the last 15 years (see, e.g., \cite{VPS10} and
references therein), while the systematic study of higher harmonics 
began quite recently \cite{qm11_phenix,qm11_cms,qm11_atlas,qm11_alice}.

In the present paper we investigate the ratio $R = v_4/v_2^2$ in
heavy-ion collisions at energies of the Relativistic Heavy Ion Collider
(RHIC) ($\sqrt{s} = 200${\it A}~GeV) and the Large Hadron Collider 
(LHC) ($\sqrt{s} = 2.76${\it A}~TeV). Interest in the study was raised 
due to the obvious discrepancy between the theoretical estimates and 
the experimental measurements. On the one hand, the exact theoretical 
result for hydrodynamics provided $v_4/v_2^2 = 0.5$ for a thermal 
freeze-out distribution \cite{BO06}. On the other hand, it was found 
soon in RHIC experiments \cite{v4v2_star,v4v2_phenix} that the measured 
ratio $R$ exceeded by factor 2 the theoretically predicted one. Both 
the STAR and the PHENIX Collaborations have reported that the $R$ is 
rather close to unity for all identified particles in a broad ranges of 
centrality, $10\% \leq \sigma/\sigma_{geo} \leq 70\%$, and transverse 
momentum, $p_T \geq 0.5$\,GeV/$c$. For the smaller $p_T$ the ratio 
seems to exceed the value of 1. Note also that the PHENIX data are 
about 10$-$15\% below the STAR ones.

In Ref.~\cite{GO10} it was argued that the experimentally measured $R$
can be larger than 0.5 even if the ratio $v_4/v_2^2$ was exactly equal
to 0.5 in each event. Such a distortion can be caused by event-by-event
fluctuations. Namely, if the ratio $v_4/v_2^2$ is estimated not on an
event-by-event basis but rather on averaging of both $v_2$ and $v_4$
over the whole statistics, the event-by-event fluctuations will 
significantly increase the extracted value of the ratio. Calculations
of $R$ at RHIC energies within both ideal and viscous hydrodynamics 
with different initial conditions \cite{LGO10} revealed that the ideal 
hydrodynamics provided better agreement with the data, although the 
STAR results remained underpredicted a bit. For LHC the hydrodynamic 
calculations have predicted similar behavior with slight increase at 
small transverse momenta \cite{LGO10}. 
  
The preliminary results obtained in Pb + Pb collisions at $\sqrt{s} =
2.76${\it A}~TeV favor further increase of the $v_4/v_2^2$ ratio
\cite{qm11_cms,qm11_atlas}. Moreover, this ratio is not a constant at 
$p_T \geq 0.5$\,GeV/$c$ but increases with rising transverse momentum. 
The first aim of the present paper is to study to what extent the hard 
processes, i.e., jets, can affect the ratio $R$ predicted by the 
hydrodynamic calculations. 

The second aim of the paper is investigation of the fulfillment of 
the so-called number-of-constituent-quark (NCQ) scaling, observed 
initially for the partial elliptic of mesons and baryons at RHIC 
\cite{ncq_star,ncq_phen}. Despite the general expectations, the 
measurements show that the NCQ scaling is broken at LHC energies
\cite{ncq_alice}. Thus, it would be interesting to elucidate the role 
of jets in the scaling violation. For these purposes we employ the 
{\small HYDJET}++ model \cite{hydjet++}, which couples the 
parametrized hydrodynamics to jets. The soft part of the {\small 
HYDJET}++ simulated event represents the thermalized hadronic state 
where particle multiplicities are determined under assumption of 
thermal equilibrium. Hadrons are produced on the hypersurface, 
represented by a parametrization of relativistic hydrodynamics with 
given freeze-out conditions. At the freeze-out stage the system breaks 
up into hadrons and their resonances. The table of baryon and meson 
resonances implemented in the model is quite extensive. This allows 
for better accounting of the influence of final-state interactions on 
the generated spectra. The hard part of the model accounts for jet 
quenching effect, i.e., radiation and collisional losses of partons 
traversing hot and dense media. The contribution of soft and hard 
processes to the total multiplicity of secondaries depends on both 
centrality of the collision and its energy and is tuned by model 
parameters to RHIC and LHC data.

The paper is organized as follows. A brief description of the {\small
HYDJET}++ is given in Sec.~\ref{sec2}. Section~\ref{sec3} presents the 
results of calculations of both $v_2$ and $v_4$ for charged particles 
in both considered reactions. The even components of the anisotropic 
flow and their ratio $R = v_4/v_2^2$ are studied in the interval $10\% 
\leq \sigma/\sigma_{geo} \leq 50\%$ in four centrality bins. In 
Sec.~\ref{sec4} the interplay between jets and decays of resonances, 
as well as  the roles of resonance decays in better realization and 
the jets in violation of the number-of-constituent-quark scaling are 
discussed. Conclusions are drawn in Sec.~\ref{sec5}.

\section{The HYDJET++ event generator}
\label{sec2}

The Monte Carlo event generator {\small HYDJET}++ \cite{hydjet++} was 
developed for fast but realistic simulation of hadron spectra in both 
central and non-central heavy-ion collisions at ultrarelativistic 
energies. It consists of two parts. The {\small FASTMC} 
\cite{fastmc1,fastmc2} event generator deals with the hydrodynamic 
evolution of the fireball. Therefore, it describes the soft parts of 
particle spectra with the transverse momenta $p_T \leq 2$\,GeV/$c$. 
The hard processes are simulated by the {\small HYDJET} model 
\cite{hydjet} that propagates jets through hot and dense partonic 
medium. Both parts of the {\small HYDJET}++ generate particles 
independently. 

To allow for really fast generation of the spectra the {\small FASTMC} 
employs a parametrized hydrodynamics with Bjorken-like or Hubble-like 
freeze-out surface parametrization. Since at ultrarelativistic 
energies the particle densities at the stage of chemical freeze-out 
are quite high, a separation of the chemical and thermal freeze-out is
also implemented. The mean number of participating nucleons $N_{part}$ 
at a given impact parameter $b$ is calculated from the Glauber model 
of independent inelastic nucleon-nucleon collisions. After that the 
value of effective volume of the fireball $V_{eff}$, that is directly 
proportional to $N_{part}$, is generated. When the effective volume of 
the source is known, the mean multiplicity of secondaries produced at 
the spacelike freeze-out hypersurface is calculated. Parametrizations
of the odd harmonics of the anisotropic flow are not implemented in
the present version of {\small HYDJET}++, whereas the elliptic flow is 
generated by means of the hydro-inspired parametrization that depends 
on momentum and spatial anisotropy of the emitting source. The model 
utilizes a very extensive table of ca. 360 baryon and meson resonances 
and their antistates together with the decay modes and branching 
ratios taken from the {\small SHARE} particle decay table \cite{share}. 
After the proper tuning of the free parameters, the {\small HYDJET}++ 
simultaneously reproduces the main characteristics of heavy-ion 
collisions at RHIC and at LHC, such as hadron spectra and ratios, 
radial and elliptic flow, and femtoscopic momentum correlations.

The multiple scattering of hard partons in the quark-gluon plasma (QGP) 
is generated by means of the {\small HYDJET} model. This approach takes
into account accumulating energy loss, the gluon radiation, and 
collisional loss, experienced by a parton traversing the QGP. The 
shadowing effect \cite{Tyw_07} is implemented in the model as well.
The {\small PYQUEN} routine \cite{pyquen} generates a single hard $NN$ 
collision. The simulation procedure includes the generation of the 
initial parton spectra with {\small PYTHIA} \cite{pythia} and 
production vertexes at a given impact parameter, 
rescattering-by-rescattering simulation of the parton path length in a 
dense medium, radiative and collisional energy losses, and final 
hadronization for hard partons and in-medium emitted gluons according 
to the Lund string model \cite{lund}. Then, the full hard part of the 
event includes {\small PYQUEN} multi-jets generated around its mean 
value according to the binomial distribution. The mean number of jets 
produced in {\it A + A} events is a product of the number of binary 
$NN$ sub-collisions at a given impact parameter and the integral 
cross section of the hard process in $NN$ collisions with the minimal 
transverse momentum transfer, $p_T^{\rm min}$. Further details of the 
model can be found in Refs.~\cite{hydjet++,fastmc1,fastmc2,hydjet}. 

It is worth mentioning recent important modification of the {\small
HYDJET}++. After the measurement of particle spectra in $pp$ 
collisions at LHC it became clear that the set of model parameters 
employed by the {\small PYTHIA}~6.4 version had to be tuned. Several 
modifications have been proposed \cite{p_perugia,p_atlas}. The 
application of standard {\small PYTHIA}~6.4 in the {\small HYDJET}++ 
led to too early suppression of elliptic flow of charged particles at 
intermediate transverse momenta in lead-lead collisions and, 
therefore, to the prediction of a weaker $v_2$ \cite{v2_prc09,sqm09} 
compared to the data. Recently, the {\small HYDJET}++ was modified 
\cite{hydjet_12} to implement the {\small Pro-Q20} tune of PYTHIA. 
In contrast to calculations of elliptic flow presented in 
\cite{v2_prc09,sqm09,sqm11}, all simulations of Pb + Pb reactions at 
LHC energies in the present paper are performed with the upgraded 
{\small HYDJET}++.

\section{$v_2$ and $v_4$ from hydrodynamics and from jets}
\label{sec3}

\begin{figure}
\resizebox{\linewidth}{!}{
\includegraphics[scale=0.60]{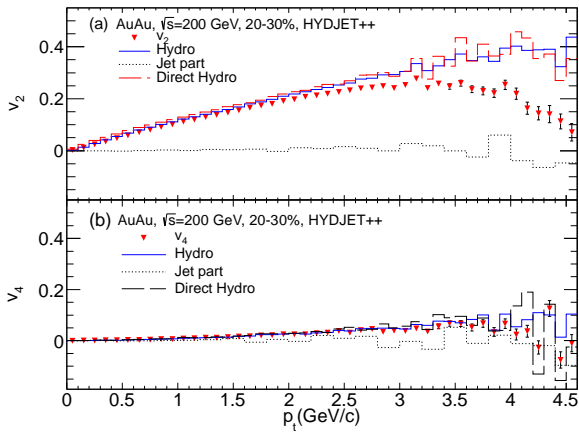}
}
\caption{(Color online)
Transverse momentum dependencies (triangles) of (a) $v_2$ and 
(b) $v_4$ of charged hadrons calculated within the {\small HYDJET}++ 
for Au + Au collisions at $\sqrt{s} = 200${\it A}~GeV at centrality 
$\sigma/\sigma_{\rm geo} = 20 - 30\%$. Histograms show flow of 
directly produced particles in hydro-calculations (dashed lines),
total hydrodynamic flow (solid lines), and flow produced by jets
(dotted lines). 
\label{fig1} }
\end{figure}

\begin{figure}
\resizebox{\linewidth}{!}{
\includegraphics[scale=0.60]{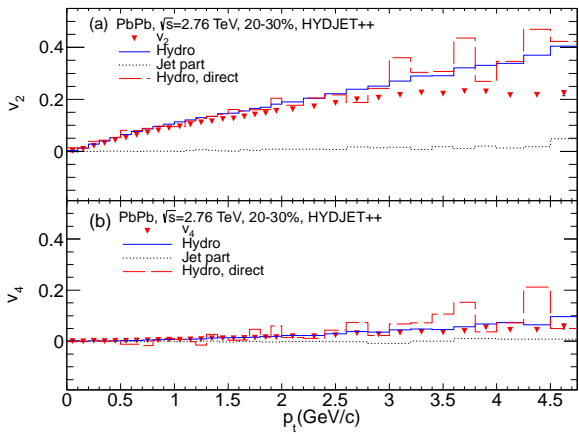}
}
\caption{(Color online)
The same as Fig.~\protect\ref{fig1} but for Pb + Pb collisions at
$\sqrt{s} = 2.76${\it A}~GeV. 
\label{fig2} }
\end{figure}

For the investigations of the second and the fourth flow harmonics, 
ca. 60 000 gold-gold and ca. 50 000 lead-lead minimum bias collisions
have been generated at $\sqrt{s} = 200${\it A}~GeV and $\sqrt{s} =
2.76${\it A}~TeV, respectively. The transverse momentum dependencies
of $v_2$ and $v_4$ obtained for the centralities 20$-$30\% are shown 
in Fig.~\ref{fig1} for RHIC and in Fig.~\ref{fig2} for LHC energies. 

Together with the resulting distributions for $v_2(p_T)$ and $v_4(p_T)$ 
we present separate contributions coming from (i) hadrons directly
produced at the freeze-out hypersurface in the hydrodynamic part, (ii)
direct and secondary hadrons created after the decays of resonances, 
and (iii) hadrons produced in the course of jet fragmentation. Recall
briefly the main features of the $v_2(p_T)$ behavior in {\small 
HYDJET}++. The elliptic flow rises up to its maximum at intermediate 
$p_T$ around 2.5$-$3\,GeV/$c$ and then rapidly drops. This falloff is 
observed in experimental data also. In the model its origin is traced 
to the interplay between the soft hydrolike processes and hard jets, as 
was studied in details in \cite{v2_prc09,sqm09}. The ideal 
hydrodynamics demonstrates continuous increase of the elliptic flow 
with rising transverse momentum. Because of the jet quenching the jets 
also develop an asimuthal anisotropy that increases with the $p_T$ too; 
however, this effect is quite weak and does not exceed few percent. 
The particle yield as a function of the transverse momentum drops 
more rapidly for hydroproduced hadrons than for hadrons from jets. 
Therefore, after a certain $p_T$ threshold jet particles start to 
dominate the particle spectrum, thus leading to a weakening of the 
combined elliptic flow. A similar tendency is observed in 
Fig.~\ref{fig1} and Fig.~\ref{fig2} for the $v_4$ also, but, because 
of the quite weak signal in the hydrodynamic part, the effect of the 
$v_4$ falloff is not as pronounced as that of the elliptic flow.  

As shown in Fig.~\ref{fig1} decays of resonances can change the 
elliptic flow of directly produced hadrons with $p_T \leq 3$\,GeV/$c$
by 1$-$2\% at RHIC and by less than 1\% at LHC; see Fig.~\ref{fig2}. 
For the $v_4$ the difference between the two histograms is negligible; 
i.e., resonance decays play a minor role for soft parts of both 
$v_2(p_T)$ and $v_4(p_T)$ distributions. At $p_T \approx 2.5$\,GeV/$c$ 
jets come into play and change dramatically the shapes of the elliptic 
and hexadecapole flows.

\begin{figure}
\resizebox{\linewidth}{!}{
\includegraphics[scale=0.65]{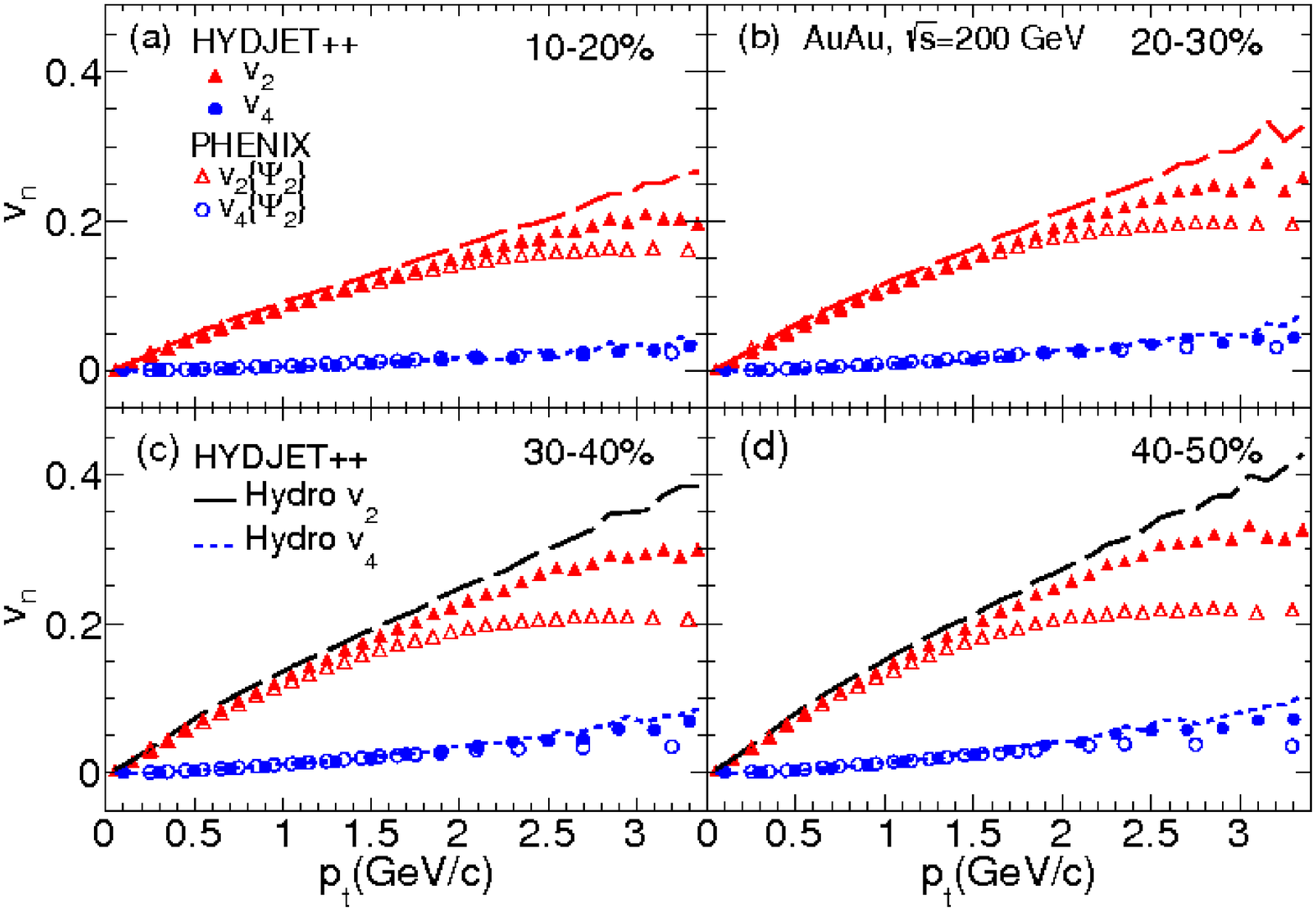}
}
\caption{(Color online) 
$v_2(p_T)$ (full triangles) and $v_4(p_T)$ (full circles) for charged 
particles in {\small HYDJET}++ calculations of Au + Au collisions at 
$\sqrt{s} = 200${\it A}~GeV at centrality $\sigma/\sigma_{\rm geo}$ 
(a) $10--20\%$, (b) $20--30\%$, (c) $30--40\%$ and (d) $40--50\%$,
respectively. Dashed lines show hydrodynamic part of the calculations.
Data from \cite{v4v2_phenix} are shown by open triangles ($v_2$) and
open squares ($v_4$).
\label{fig3} }
\end{figure}

\begin{figure}
\resizebox{\linewidth}{!}{
\includegraphics[scale=0.65]{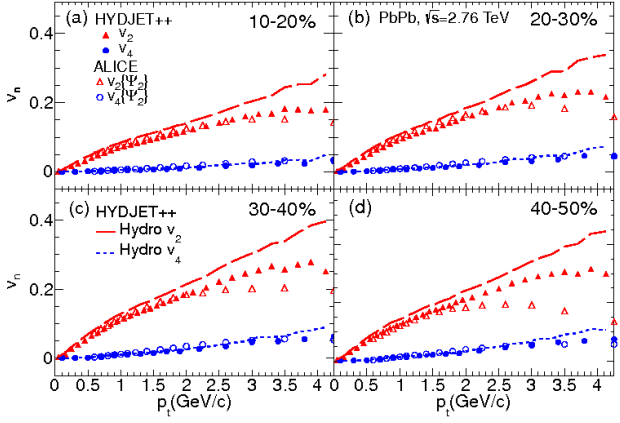}
}
\caption{(Color online) 
The same as Fig.\protect\ref{fig3} but for Pb + Pb collisions at
$\sqrt{s} = 2.76${\it A}~GeV. Experimental data are taken from
\cite{alice_flow}. 
\label{fig4} }
\end{figure}

It is worth discussing here details concerning the determination of the
flow components in the experiment and in the model. In the HYDJET++ 
simulations the elliptic flow is connected to the eccentricity of
overlapped volume of colliding nuclei. No fluctuations in the location
of nucleons within the overlapped zone are considered. Therefore, the 
flow is determined with respect to the position of true reaction plane.
The next even component, $v_4$, is not parametrized in the present 
version of the model; i.e., the hexadecapole flow comes out here merely
due to the elliptic flow. Thus, it should also be settled by the 
position of the true reaction plane. Because of the absence of the
fluctuations and non-flow effects, the ratio $v_4/v_2^2$ obtained on
an event-by-event basis equals that extracted by separate averaging
of $v_4$ and $v_2$ over the whole simulated statistics.

In the experiment the situation is more complex. For instance, in the
standard event plane (EP) method the event flow vector $\vec{Q_n}$ for
$n$-th harmonic is defined as (see \cite{VPS10} for details)
\beqar
\ds
\nonumber
\vec{Q_n} &=& (Q_{n,x} , Q_{n,y}) = \left( \sum \limits ^{}_{i} w_i
\cos{(n \phi_i)} , \sum \limits ^{}_{i} w_i \sin{(n \phi_i)} \right) 
\\   
\label{eq2}
      &=& \left( Q_n \cos{(n \Psi_n)} , Q_n \sin{(n \Psi_n)} \right).
\eeqar
The quantities $w_i$ and $\phi_i$ are the weight and the azimuthal 
angle in the laboratory frame for the $i$th particle, respectively. 
From Eq.~(\ref{eq2}) it follows that the event plane angle $\Psi_n$ 
can be expressed via the {\it arctan2} function, which takes into 
account the signs of both vector components to place the angle in the 
correct quadrant,
\beq
\ds
\Psi_n = \arctan2(Q_{n,y} , Q_{n,x})/n \ .
\label{eq3}
\eeq
The $n$th harmonic $v_n$ of the anisotropic flow at given rapidity 
$y$, transverse momentum $p_T$, and centrality $\sigma/\sigma_{geo}$ 
is determined with respect to the $\Psi_n$ angle
\beq
\ds
v_n(y, p_T, \sigma/\sigma_{geo}) = \langle \cos{[n(\phi_i - \Psi_n)]}
\rangle
\label{eq4}
\eeq
by averaging $\langle \ldots \rangle$ over all particles in all 
measured events. It is easy to see that the event plane angle for the
elliptic flow $\Psi_2$ does not necessarily coincide with that for the
hexadecapole flow $\Psi_4$. To compare our model results with the
experimental ones we need, therefore, the data where the fourth 
harmonic is extracted with respect to the $\Psi_2$ rather than the
$\Psi_4$ event plane angle.

To demonstrate the development of both $v_2$ and $v_4$ at different 
centralities, we display the flow harmonics for charged particles in
heavy-ion collisions at RHIC and LHC energies in Figs.~\ref{fig3} and 
\ref{fig4}, respectively. The experimental data by the PHENIX (RHIC) 
and the ALICE (LHC) Collaborations are plotted onto the simulations as 
well. One can see here that {\small HYDJET}++ overestimates the 
elliptic flow of charged hadrons with transverse momenta $2\,{\rm GeV}
/c \leq p_T \leq 4$\,GeV/$c$ in both reactions considered. This 
indicates that simplified combination of ideal hydrodynamics and jets 
is probably enough to simulate first two even harmonics of anisotropic 
flow at $p_T \leq 2$\,GeV/$c$, whereas at higher transverse momenta 
other mechanisms, e.g., coalescence, should be taken into account for 
better quantitative description of the flow behavior. 

The elliptic flow produced by the jet hadrons with $p_T \leq 2$~GeV/$c$
is almost zero. Because of the jet quenching, the flow increases to
$3--5\%$ with rising transverse momentum; however, the jets alone 
cannot provide strong flow signal, say $v_2 \approx 10\%$, even at
LHC energies. Since the $v_4$ created by jets is also very small, it
would be instructive to study how the admixture of jet hadrons can
alter the $v_4/v_2^2$ ratio.

\begin{figure}
\resizebox{\linewidth}{!}{
\includegraphics[scale=0.65]{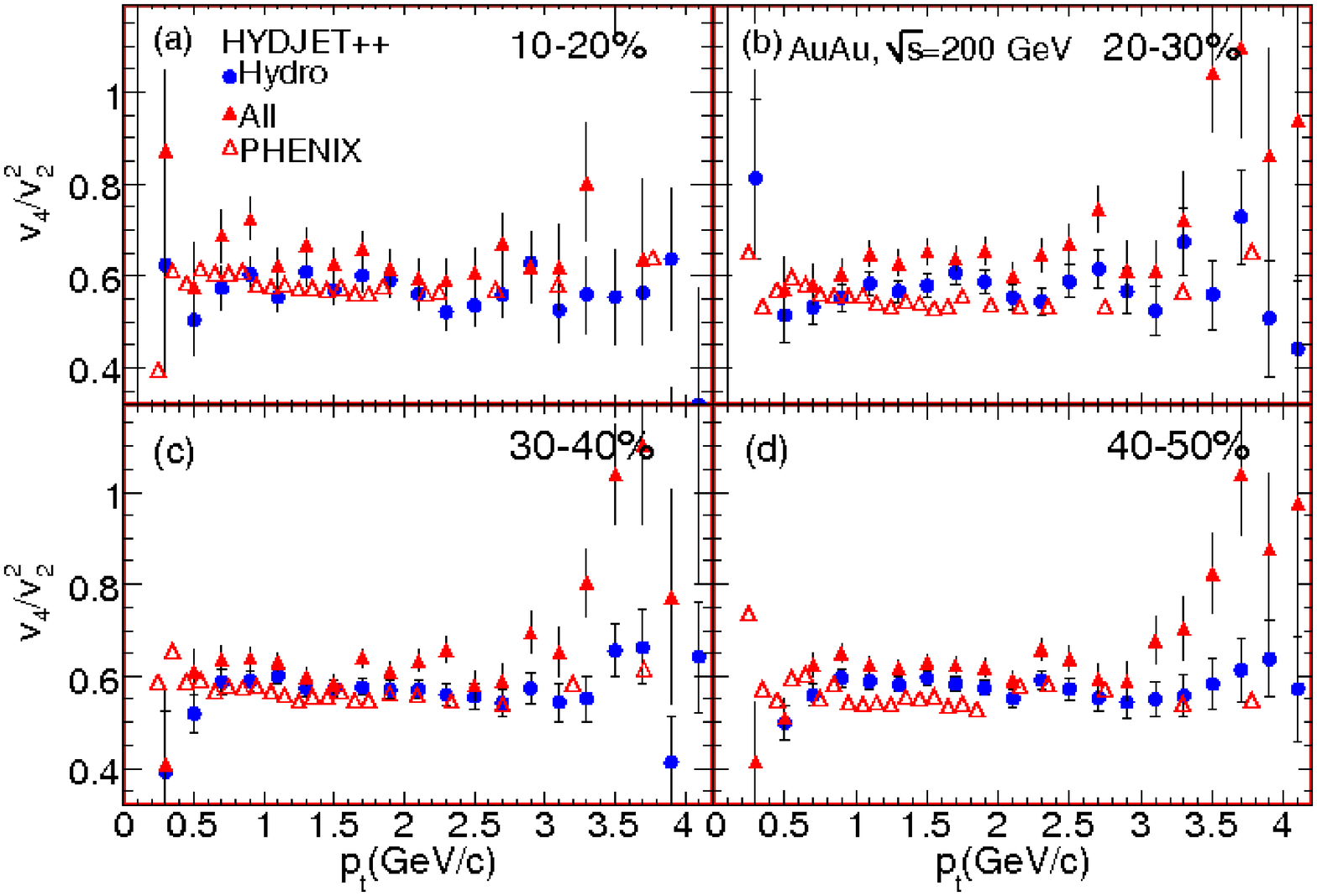}
}
\caption{(Color online) 
Ratio $v_4/(v_2)^2$ vs. $p_T$ for charged particles in {\small 
HYDJET}++ calculations of Au + Au collisions at $\sqrt{s} = 
200${\it A}~GeV at centrality $\sigma/\sigma_{\rm geo}$ (a) $10--20\%$, 
(b) $20--30\%$, (c) $30--40\%$ and (d) $40--50\%$, respectively. Full 
circles denote the hydro+jet calculations, open circles show only 
hydro-part, and open squares indicate the rescaled experimental data 
(see text for details).
\label{fig5} }
\end{figure}

\begin{figure}
\resizebox{\linewidth}{!}{
\includegraphics[scale=0.65]{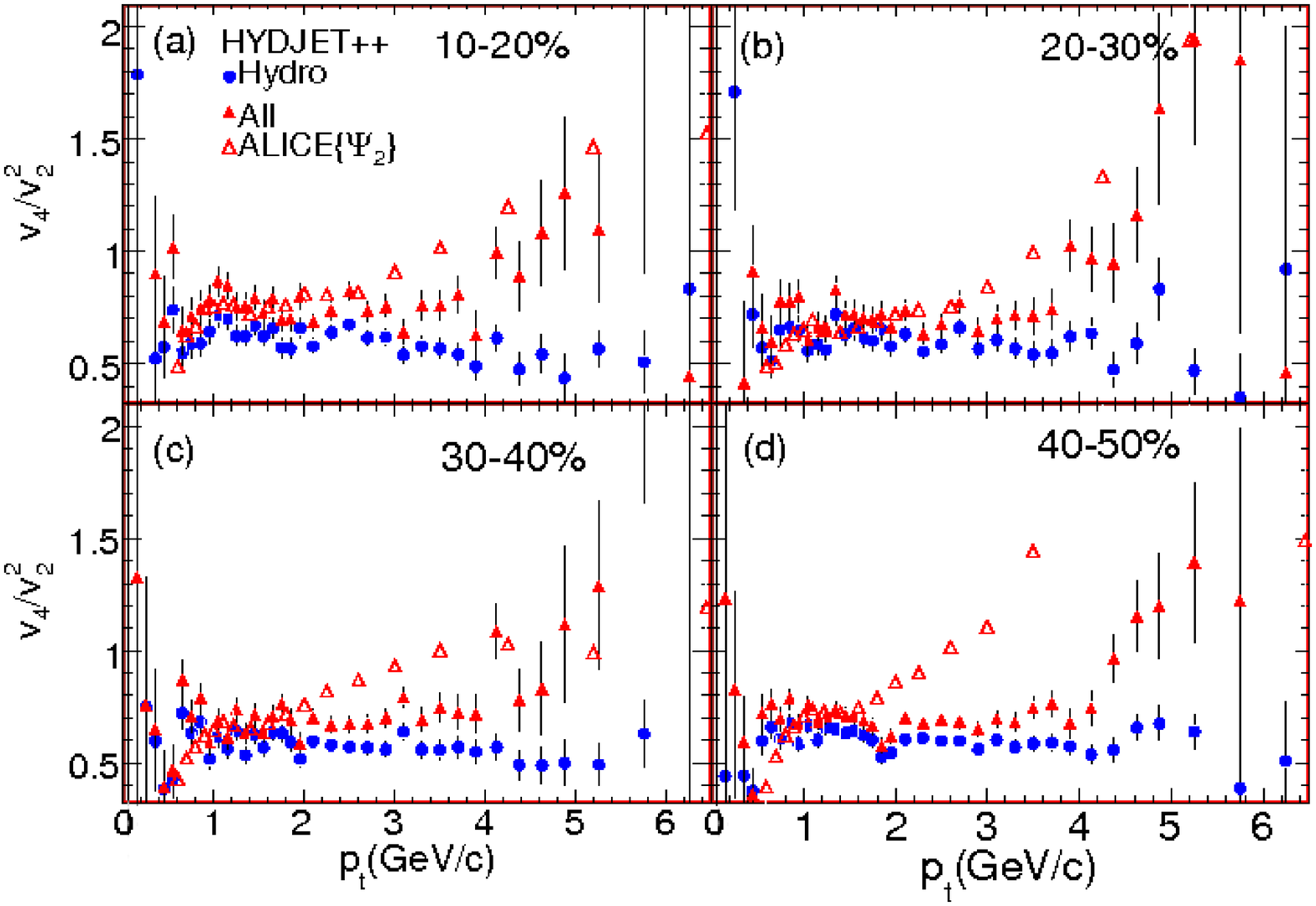}
}
\caption{(Color online) 
The same as Fig.~\protect\ref{fig5} but for Pb + Pb collisions at
$\sqrt{s} = 2.76${\it A}~GeV. 
\label{fig6} }
\end{figure}

The ratio $R = v_4/v_2^2$ as a function of transverse momentum is 
presented in Figs.~\ref{fig5} and \ref{fig6} for four different
centralities in Au + Au collisions at RHIC and in Pb + Pb collisions 
at LHC, respectively. The final result is compared here to the ratio
obtained merely for hydro-like processes and to the experimental data. 
As was mentioned in \cite{GO10}, the measured ratio should be 
noticeably larger than 0.5. There are event-by-event fluctuations that 
increase $R$ even if both flow harmonics are determined by means of 
the $\Psi_2$ event plane angle. The increase occurs because of the 
averaging of both $v_2$ and $v_4$ over the whole event sample before 
taking the ratio. These fluctuations are lacking in the {\small 
HYDJET}++; therefore, the data used for the comparison are properly 
reduced. See \cite{GO10,LGO10} for details. It is seen that 
parametrized hydrodynamics with the extended table of resonances 
already provides $v_4/v_2^2 \approx 0.6$, which is higher than the 
theoretical value of $R = 0.5$. Jet particles increase this ratio 
further to value $R \approx 0.65$ at RHIC and $R \approx 0.7$ at LHC. 
While the ratio $R$ is insensitive to the transverse momentum at $0.1 
{\rm GeV}/c \leq p_T \leq 3$\,GeV/$c$, at higher $p_T$ it increases 
with rising transverse momentum both in model simulations and in the 
experiment, although the RHIC data favor a weaker dependence. Thorough 
study of this problem within the hydrodynamic model indicates 
\cite{LGO10} that neither the initial conditions nor the shear 
viscosity can be accounted for the rise of high-$p_T$ tail of the 
distribution. It looks like this rise can be attributed solely to jet 
phenomenon.

At LHC energy the increase of $R$ with rising transverse momentum at 
$p_T \geq 3$\,GeV/$c$ is quite distinct. The difference between the
model results and the data visible for semiperipheral collisions at
$40\% \leq \sigma/\sigma_{geo} \leq 50\%$ can be partly explained by 
the imperfect description of the elliptic flow at $p_T \geq 
2.5$\,GeV/$c$; see Fig.~\ref{fig4}. Also, the STAR results concerning
the $v_2$ are about 15$--$20\% higher than the PHENIX data, and the 
{\small HYDJET}++ model is tuned to averaged values provided by these 
two RHIC experiments. Nevertheless, the effect of hard processes is 
clear: The hydrodynamic part of the code yields rather flat ratio 
$v_4/v_2^2$, whereas the jets provide the rise of the high-$p_T$ tail.

\section{Number-of-constituent-quark scaling}
\label{sec4}

The number-of-constituent-quark (NCQ) scaling in the development of
elliptic flow was first observed in Au + Au collisions at RHIC
\cite{ncq_star,ncq_phen}. If the elliptic flow, $v_2$, and the
transverse kinetic energy, $K E_T \equiv m_T - m_0$, of any hadron
species are divided by the number of constituent quarks, i.e.,
$n_q = 3$ for a baryon and $n_q = 2$ for a meson, then the scaling
in $v_2(K E_T)$ holds up until $K E_T/n_q \approx 1$\,GeV 
\cite{PHENIX}. The observation of the NCQ scaling seems to favor the 
idea of the elliptic flow formation already on a partonic level. For
instance, as pointed out in \cite{ncq_break}, the scaling is broken 
if hadrons are produced in the course of string fragmentation, 
whereas the process of quark coalescence leads to the scaling 
emergence. On the other hand, as was shown in 
Refs.~\cite{v2_prc09,sqm09}, the fulfillment of the NCQ scaling at 
ultrarelativistic energies depends strongly on the interplay between 
the decays of resonances and jets. Note that the breaking of the NCQ 
scaling at LHC was observed experimentally in 
\cite{ncq_alice,ncq_alice2}.

\begin{figure}
\resizebox{\linewidth}{!}{
\includegraphics[width=0.75\textwidth]{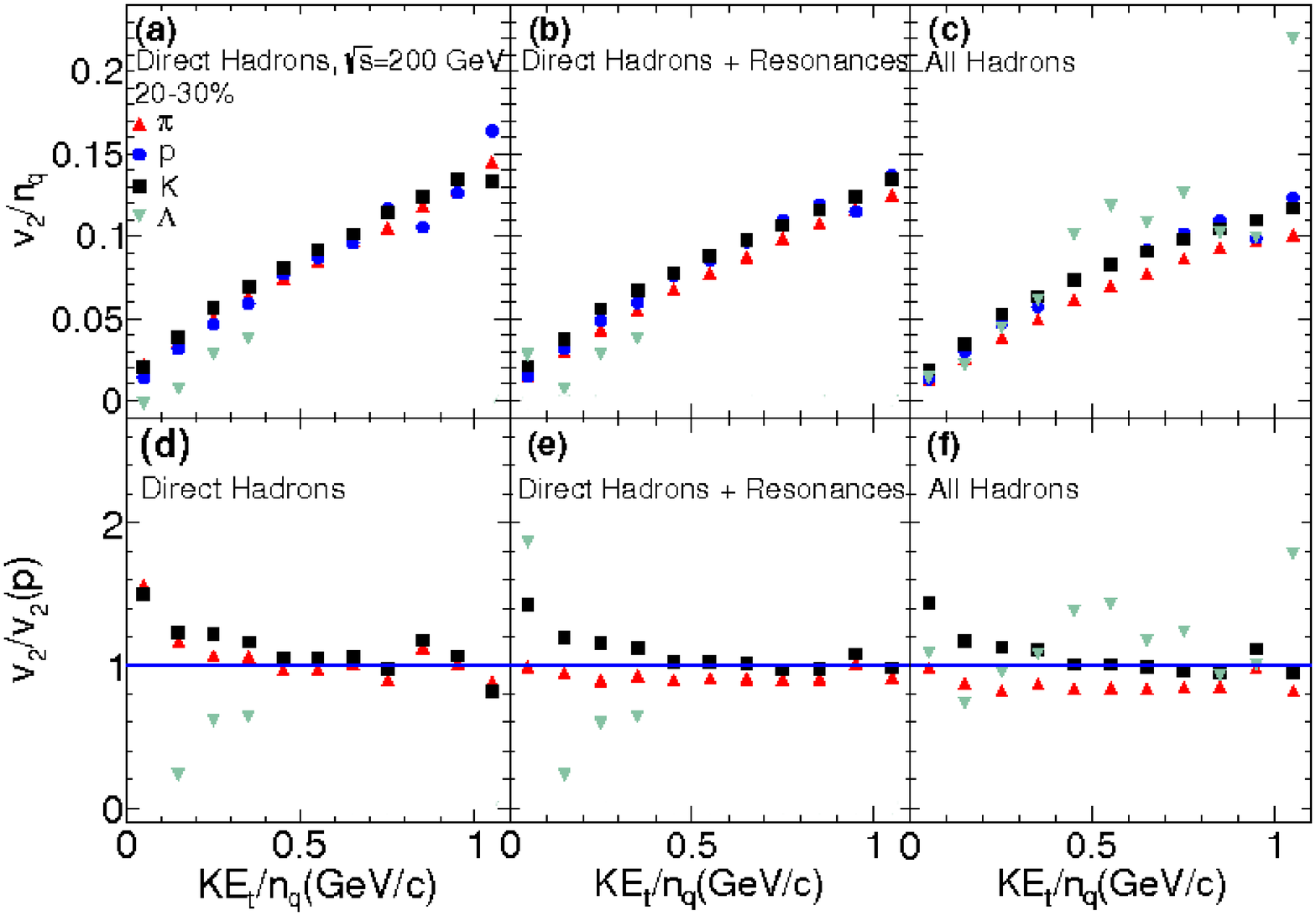}
}
\caption{(Color online) Upper row: The $KE_T/n_q$ dependence of 
elliptic flow for (a) direct hadrons, (b) hadrons produced both 
directly and from resonance decays, and (c) all hadrons produced in 
the {\small HYDJET}++ model for Au + Au collisions at $\sqrt{s} = 
200${\it A}~GeV with centrality 20$--$30\%. Bottom row: The 
$KE_T/n_q$ dependence of the ratios $(v_2/n_q)\left/(v_2^p/3) 
\right.$ for (d) direct hadrons, (e) direct hadrons plus hadrons 
from the decays, and (f) all hadrons.
\label{fig7} }
\end{figure}

\begin{figure}
\resizebox{\linewidth}{!}{
\includegraphics[width=0.75\textwidth]{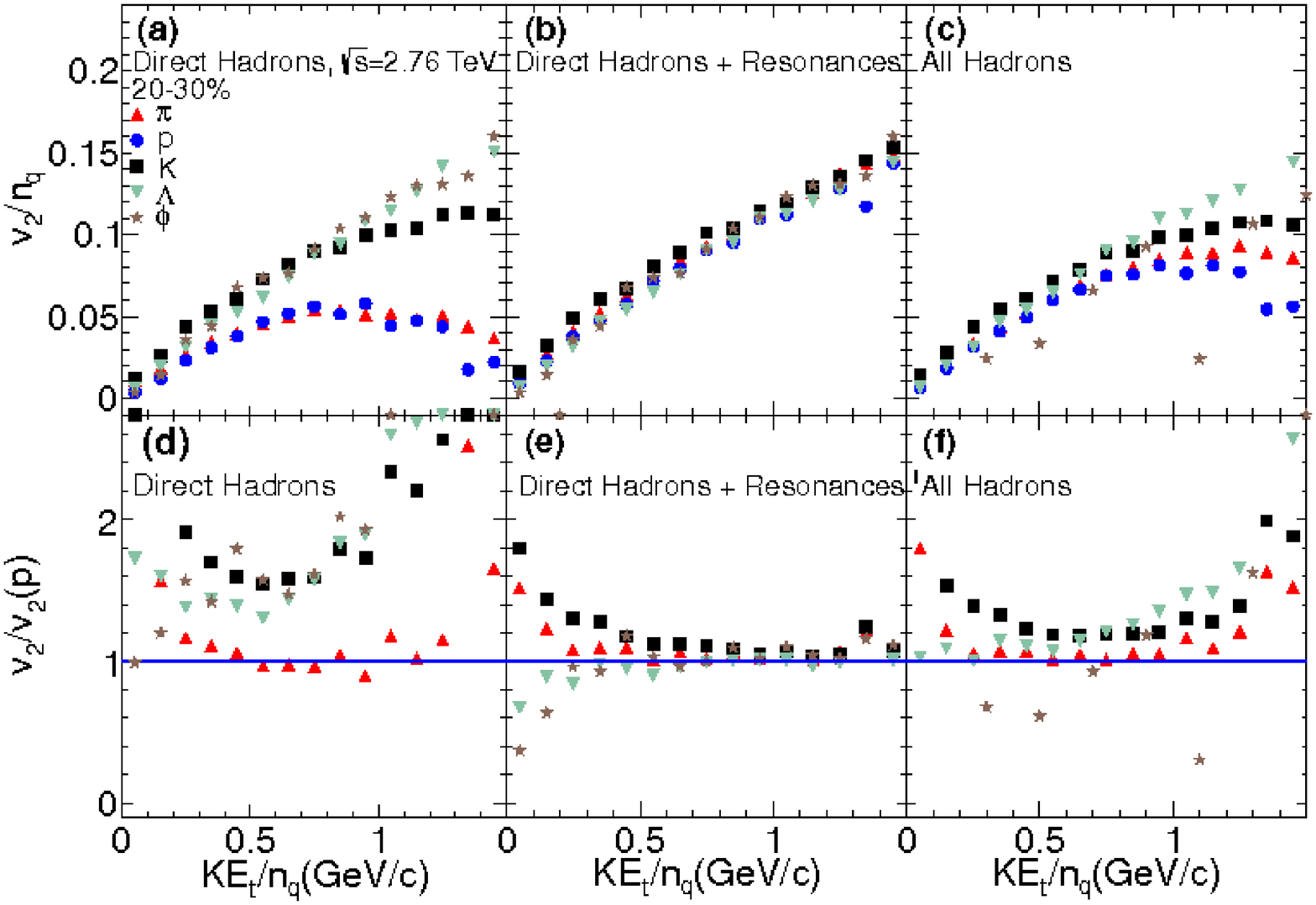}
}
\caption{(Color online) The same as Fig~.\protect\ref{fig7} 
but for Pb + Pb collisions at $\sqrt{s}=2.76${\it A}~TeV.
\label{fig8} }
\end{figure}

To demonstrate the importance of both resonance decays and jets for 
the formation of NCQ scaling we plot the reduced functions 
$v_2^h/n_q (KE_T/n_q)$ for several hadronic species obtained in 
{\small HYDJET}++ simulations of heavy-ion collisions at RHIC 
(Fig.~\ref{fig7}) and at LHC (Fig.~\ref{fig8}) energies in 
centrality bin 20$--$30\%. These distributions are then also 
normalized to the flow of protons, $v_2^h/n_q : v_2^p/3$, to see 
explicitly degree of the scaling fulfillment. The study is subdivided 
into three steps. The flow of hadrons straight after the thermal 
freeze-out in hydrodynamic calculations is displayed in left windows. 
Central windows present this flow modified by the final state 
interactions, i.e., decays of resonances. Finally, right windows show 
the resulting flow of hadrons coming from all processes. 

At RHIC energy, it looks like at given centrality the direct pions, 
protons and kaons are produced already obeying the scaling within the 
5$--$10\% accuracy limit; see Figs.~\ref{fig7}(a) and \ref{fig7}(d). 
The scaling holds also after decays of resonances as demonstrated in 
Figs.~\ref{fig7}(b) and \ref{fig7}(e). Its fulfillment becomes 
slightly worse when hadrons from jets are taken into account; however, 
the NCQ scaling remains valid within 10\% accuracy at least for three 
main hadron species. The situation is drastically changed for the 
collisions at LHC. Here spectra of directly produced particles do not 
possess any scaling properties, as one can see in Figs.~\ref{fig8}(a) 
and \ref{fig8}(d). After final-state interactions the scaling 
conditions for hadrons in hydrodynamic simulations are restored, as 
displayed in Figs.~\ref{fig8}(b) and \ref{fig8}(e). Even $\phi$ mesons 
follow the unique trend. Why? Spectra of many light hadrons, especially 
pions and protons, are getting feed-down from heavy resonances, whereas 
the spectrum of $\phi$ remains unchanged. The resonance boost makes 
elliptic flows of light hadrons harder. As a result, the NCQ scaling is 
fulfilled in a broad range of $KE_T/n_q$ in the hydro sector of the 
model. In contrast, hard processes cause significant distortions of 
particle spectra and lead to violation of the scaling conditions; see 
Figs.~\ref{fig8}(c) and \ref{fig8}(f), in accordance with experimental 
observations \cite{ncq_alice,ncq_alice2}.

\section{Conclusions}
\label{sec5}

Formation of elliptic $v_2$ and hexadecapole $v_4$ flows of hadrons in
Au + Au collisions at $\sqrt{s} = 200${\it A}~GeV and in Pb + Pb 
collisions at $\sqrt{s} = 2.76${\it A}~TeV is studied within the 
{\small HYDJET}++ model. This model combines the parametrized 
hydrodynamics with hard processes (jets). Therefore, the main aim was 
to investigate the role of interplay between soft and hard processes 
for the development of flow. Several features have been observed. 
First, the jets are found to increase the ratio $R = v_4 / v_2^2$ for 
both considered heavy-ion reactions. Second, jets lead to rise of the 
high-$p_T$ tail of the ratio $R$. Such a behavior is observed 
experimentally but cannot be reproduced by conventional hydro models 
relying on ideal or viscous hydrodynamics. Third, the resonance 
feed-down significantly enhances the flow of light hadrons and modifies 
their spectra toward the fulfillment of number-of-constituent-quark 
scaling. The flow of particles produced in jet fragmentation is quite 
weak, thus jets are working against the scaling. Due to interplay of 
resonance and jet contribution, the NCQ scaling works well only at 
certain energies, where jets are not abundant. Because jet influence 
increases with rising collision energy, just approximate NCQ scaling 
is observed at LHC despite the fact that the scaling holds for the 
pure hydrodynamic part of hadron spectra. At higher collision energies 
scaling performance should get worse. 

\begin{acknowledgments}
Fruitful discussions with I.~Lokhtin, L.~Malinina, I.~Mishustin, and 
A.~Snigirev are gratefully acknowledged. We are thankful to 
J.-Y.~Ollitrault and K.~Redlich for bringing to our attention the 
$v_4/v_2^2$ problem. 
This work was supported in part by the QUOTA Program and Norwegian 
Research Council (NFR) under Contract No. 185664/V30.
\end{acknowledgments}


\end{document}